\documentclass[9pt,twocolumn,twoside]{osajnl}

\journal{ol} 

\newcommand{\sj}[6]{ \begin{Bmatrix}
  #1 & #2 & #3 \\
  #4 & #5 & #6
 \end{Bmatrix}}

\newcommand{\new}[1]{\textcolor{black}{#1}}
\newcommand{\old}[1]{} 

\newcommand{\comm}[1]{} 
\newcommand{\LQ}[1]{\textrm{#1}} 

\setboolean{shortarticle}{true}

\ifthenelse{\boolean{shortarticle}}{\colorlet{color2}{color2b}}{\colorlet{color2}{color2}} 

\title{Doppler-free approach to optical pumping dynamics in the $6\LQ{S}_{1/2}- 5\LQ{D}_{5/2}$ electric quadrupole transition of Cesium vapor}

\author[1]{Eng Aik Chan}
\author[1]{Syed Abdullah Aljunid}
\author[1,2]{Nikolay I. Zheludev}
\author[1,3,4,5,*]{David Wilkowski}
\author[1,5,6,+]{Martial Ducloy}

\affil[1]{Centre for Disruptive Photonic Technologies \& Division of Physics and Applied Physics, SPMS, Nanyang Technological University, 637371, Singapore.}
\affil[2]{Optoelectronics Research Centre \& Centre for Photonic Metamaterials, University of Southampton, Southampton SO17}
\affil[3]{Centre for Quantum Technologies, National University of Singapore, 117543 Singapore}
\affil[4]{MajuLab, CNRS-UNS-NUS-NTU International Joint Research Unit UMI 3654, Singapore}
\affil[5]{School of Physical and Mathematical Sciences, Nanyang Technological University, 637371 Singapore, Singapore}
\affil[6]{Laboratoire de Physique des Lasers, Universit\'{e} Paris 13, CNRS, (UMR 7538), F-93430, Villetaneuse, France.}

\affil[*]{Electronic address: david.wilkowski@ntu.edu.esg}
\affil[+]{Corresponding author: martial.ducloy@ntu.edu.sg}

\dates{}

\ociscodes{(020.2930) Hyperfine structure; (300.6420) Spectroscopy, nonlinear; (300.6210) Spectroscopy, atomic; (140.0140) Lasers and laser optics}

\doi{\url{http://dx.doi.org/10.1364/OL.41.002005}}

\begin{abstract}
The $6\LQ{S}_{1/2}-5\LQ{D}_{5/2}$ electric quadrupole transition is investigated in Cesium vapor at room temperature via nonlinear Doppler-free $6\LQ{P}$-$6\LQ{S}$-$5\LQ{D}$ three-level spectroscopy.
Frequency-resolved studies of individual E2 hyperfine lines allow one to analyze optical pumping dynamics, polarization selection rules and line intensities.
It opens the way to studies of transfer of light orbital angular momentum to atoms, and the influence of metamaterials on E2 line spectra.
\end{abstract}

\setboolean{displaycopyright}{true}

\begin{document}

\maketitle
\thispagestyle{fancy}

\ifthenelse{\boolean{shortarticle}}{\ifthenelse{\boolean{singlecolumn}}{\abscontentformatted}{\abscontent}}{}


With the appearance of laser sources, nonlinear Doppler-free Laser Spectroscopy (DFLS) has undergone a very fast development.
It has been utilized for atomic and molecular spectral analyses, collisional studies in the vapor phase and investigation of fundamental processes~\cite{letokhov_nonlinear_1977,demtroder_laser_2003}.
Up to now, in atomic physics, DFLS has been mainly performed by using laser sources resonant for electric-dipole (E1) transitions.
Dipole-forbidden transitions, particularly electric quadrupole (E2) transitions, are important in new avenues of atomic physics for fundamental studies like parity violation~\cite{tsigutkin_observation_2009} or devising of ultra-high-accuracy optical clocks~\cite{tamm_spectroscopy_2000,oskay_single-atom_2006,rosenband_frequency_2008}.
Spectroscopic studies of E2 transitions in vapors are generally
hindered by Doppler-broadening, and in most cases averaged over the internal structure of the E2 transition (e.g.\ hyperfine multiplets)~\cite{tojo_absorption_2004,tojo_precision_2005}.
A noteworthy exception is the early work by Weber and Sansonetti~\cite{weber_accurate_1987} who performed resonantly enhanced stepwise excitation to high lying states of Cesium, using the $5\LQ{D}_{3/2}$ level as the intermediates state.
In this way, they have been able to get Doppler-free spectra and resolve the hyperfine lines of the $6\LQ{S}_{1/2}-5\LQ{D}_{3/2}$ E2 transition.
Recent studies include Doppler-free 5p-6p transitions in Rubidium~\cite{ponciano-ojeda_observation_2015} and magnetic-field-mixing of forbidden hyperfine transitions of Cs D2 line~\cite{sargsyan_giant_2014}.
Another well-explored approach to study the internal structure of highly-excited D levels of alkalis and measure their energy makes use of Doppler-free two-photon spectroscopy~\cite{nez_optical_1993,kumar_precision_2014}.
In this letter we analyze Doppler-free hyperfine spectral lines of the Cs $6\LQ{S}_{1/2}-5\LQ{D}_{5/2}$ E2 transition, as observed via three-level Raman-type nonlinear spectroscopy~\cite{feld_laser-induced_1969,ducloy_laser_1978,thoumany_optical_2009} on the $6\LQ{P}$-$6\LQ{S}$-$5\LQ{D}$ coupled system.
We investigate polarization properties and optical pumping processes responsible for the E2 spectral line intensities, demonstrating the important role played by transit time relaxation.
This work should pave the way to investigate such specific properties as transfer of non-zero e.m. orbital angular momentum to atomic systems~\cite{klimov_detecting_2009,klimov_mapping_2012}, vapor-surface physics~\cite{tojo_absorption_2004,tojo_precision_2005} and atomic gas combined with nanostructured interface~\cite{deguchi_simulation_2009,kern_strong_2012,yannopapas_giant_2015}.

\begin{figure}[htbp]
\centering
\fbox{\includegraphics[width=0.55\linewidth]{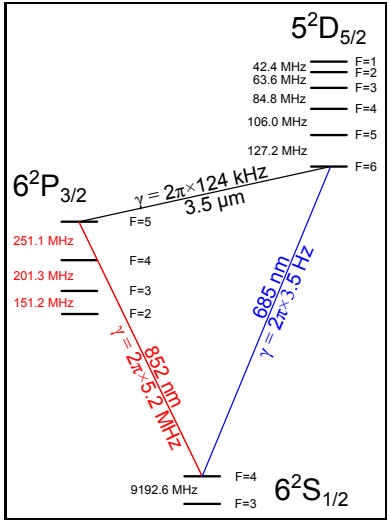}}
\caption{Cesium hyperfine states $6\LQ{S}_{1/2}$, $6\LQ{P}_{3/2}$ and $5\LQ{D}_{5/2}$ and their transitions.}
\label{fig:levels}
\end{figure}

To investigate Doppler-free spectroscopic characteristics on an E2 transition, we address the $6\LQ{S}_{1/2}\rightarrow 6\LQ{D}_{5/2}$ transition of Cesium at $\lambda = 685$\,nm (Fig.~\ref{fig:levels}).
This E2 line has a transition rate of $\gamma_{5D-6S}=2\pi \cdot 3.5$\,Hz~\cite{tojo_precision_2005}.
Considering solely fluorescence decay, the saturation intensity behaves like $I_s = \frac{2\pi^2}{3\lambda^3}\hbar c \frac{{\gamma_{5D}^{\mathrm{se}}}^2}{\gamma_{5D-6S}}\sim 2$\ Wcm$^{-2}$, where $\gamma_{5D}^{\mathrm{se}}=2\pi \cdot 124$\,kHz is the fluorescence rate of $5D_{5/2}$ state which almost exclusively comes from the $5D_{5/2}\rightarrow6P_{3/2}$ E1 line at 3.5\,µm~\cite{diberardino_lifetime_1998}.
The 685\,nm laser has an output power of 11\,mW and a minimum beam waist around $\omega_0=125$\,µm.
So the maximum laser intensity attainable is $I\approx 40$\, Wcm$^{-2}$ which should be above the saturation intensity of the transition.
However one should note that the transit time broadening, added to radiative relaxation, increases the total relaxation rate of the $5D_{5/2}$ state by more than a factor of 2, and thus $I_s$ by a factor of $\sim 8$.
To monitor Doppler-free lines at 685\,nm, we have adopted a pump-probe three-level spectroscopy approach~\cite{feld_laser-induced_1969,ducloy_laser_1978} in which population changes induced in the $6S_{1/2}$ ground state by a 685\,nm laser are monitored on the $6S_{1/2}\rightarrow 6P_{3/2}$ line transmission at $\lambda = 852$\,nm (Fig.~\ref{fig:levels}).
The radiative linewidth of the 852\,nm transition is $\gamma_{6P-6S} =2\pi \cdot 5.23$\,MHz.
As it corresponds to the fastest characteristic time of the system, this linewidth imposes the ultimate spectral resolution on the E2 transition resonances.
Taking into account the difference in wavenumbers between wavelengths at 685\,nm and at 852\,nm, we find an ultimate spectral resolution on the E2 transition resonances of $852/685 \times \gamma_{6P-6S} \approx 2\pi \cdot 6.6$\,MHz\cite{feld_laser-induced_1969,ducloy_laser_1978}.
Hence, our pump/probe spectroscopy technique would not give information on the bare linewidth of the E2 transition but it is able to resolve the hyperfine structure of the transition.

\begin{figure}[htbp]
\centering
\fbox{\includegraphics[width=0.95\linewidth]{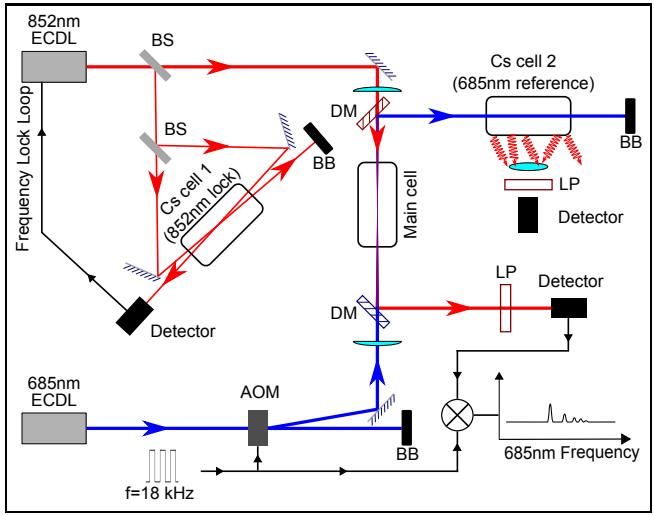}}
\caption{Experimental setup (BB: Beam Block, DM: Dichroic Mirror, LP: Long Pass filter, BS, Beam Splitter, ECDL: External Cavity Diode Laser)}
\label{fig:setup}
\end{figure}

The schematic of the experiment setup is depicted in figure~\ref{fig:setup}.
The 852\,nm External Cavity Diode Laser (ECDL) is frequency locked onto one of the hyperfine transition of the $6S_{1/2}\rightarrow 6P_{3/2}$ line, thanks to a saturated absorption side-experiment performed on an auxiliary room-temperature Cs Vapor cell (cell 1 of Fig.~\ref{fig:setup}). We also address the 685\,nm transition using an ECDL.
The 685\,nm laser beam is amplitude-modulated at 18\,kHz using an Acousto-Optic Modulator (AOM) operating at 80\,MHz.
It is focused and counter propagates with the similarly focused 852\,nm laser beam inside a 5\,cm long room-temperature main Cs vapor cell.
Inside this cell, the 852\,nm peak laser intensity is 2\,mW/cm$^2$, i.e.\ below saturation intensity. \comm{DW: Lets give the exact value of the laser intensity and updated the cover letter}
The 18\,kHz modulation amplitude, induced on the 852\,nm transmission via three-level saturation spectroscopy, is then extracted using a lock-in amplifier.
These transmission spectra are recorded as functions of the frequency of the 685\,nm laser. The 685\,nm frequency scan over the Doppler profile is monitored by collecting the fluorescence at 852\,nm in a second auxiliary vapor cell operated at temperature, $T= 55\,^\circ$C (Cell 2 in Fig. \ref{fig:setup}).
The 685\,nm frequency scale is calibrated by comparing the fluorescence spectra obtained with zeroth and first order AOM diffracted beams.

\begin{figure}[tbp]
\centering
\includegraphics[width=0.95\linewidth]{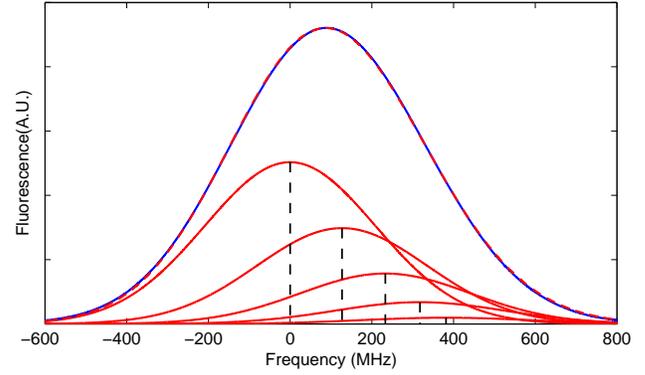}
\caption{Fluorescence spectrum at 852\,nm (dashed red curve) where only the 685\,nm laser beam is present. The solid blue curve is a fit composed of five Gaussian profiles (red curves). The dashed vertical lines correspond to the relative position of the hyperfine structure. The origin of the frequency is taken at the $F=4\rightarrow F=6$ resonance.}
\label{fig:fluo}
\end{figure}

With only the 685\,nm light, the fluorescence signal at 852\,nm comes from the radiative cascade $5D_{5/2}\rightarrow 6P_{3/2}\rightarrow 6S_{1/2}$.
An example of a fluorescence spectrum is shown in Fig.~\ref{fig:fluo}.
The 685\,nm laser is tuned from the $F=4$ hyperfine ground state.
Thus according to the selection rules of a quadrupole transition, the $F=4\rightarrow F^\prime = 2,3,4,5\, \&\, 6$ lines can be observed.
Those transitions are not resolved in Doppler spectroscopy due to the small hyperfine splitting of the $5D_{5/2}$ state.
However the asymmetry of the fluorescence spectrum is a clear signature of the presence of several transitions.
Moreover, we calculate the transitions relative intensities, based only on the E2 absorption line strengths $(J,F\rightarrow J^\prime,F^\prime)$\cite{weissbluth_atoms_1978}:
\begin{equation}
S^Q_{FF^\prime}=(2F^\prime +1)(2J+1)
\sj{J}{J^\prime}{2}{F^\prime}{F}{I}^2,
\label{eqn:E2_str}
\end{equation}
with $I=7/2$.
We now place a Gaussian profile for each hyperfine transition at its correct relative frequency separation (see Fig.~\ref{fig:fluo}) with a relative weight, given by \eqref{eqn:E2_str}.
Using a global frequency shift and an identical width for the Gaussians as the two free fitting parameters, we are able to recover the experimental profile of the fluorescence signal at 852\,nm with an excellent agreement.
\old{The Gaussian width corresponds to a temperature of, $T=47\,^\circ$C, compatible with the measured value ($T=55\,^\circ$C in cell 2).}

\begin{figure}[htbp]
\centering
\includegraphics[width=0.95\linewidth]{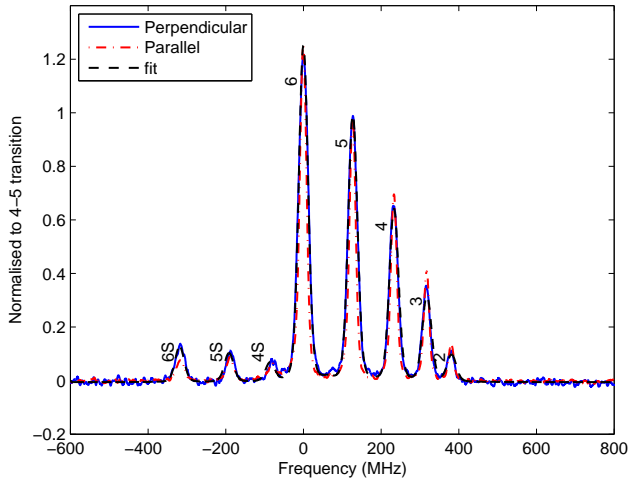}
\caption{Doppler free spectra with 852\,nm light locked on $6S_{1/2}(F=4)\rightarrow 6P_{3/2}(F^\prime=5)$ and 685\,nm light scanned across the $6S_{1/2}(F=4)$ to $5D_{5/2}(F^\prime=\mbox{2--6})$ transitions.
Peaks marked with S correspond to sideband contributions from \old{852\,nm} the $6S_{1/2}(F=4) \rightarrow 6P_{3/2}(F^\prime=4)$ transition \old{from}\new{for} a non-zero velocity group.
The blue solid curve corresponds to the spectrum where the 685\,nm and 852\,nm beams polarizations are perpendicular. The dotted dashed curve is for parallel polarizations. The dashed black curve is a fit of \eqref{eqn:signal} with $\alpha = 0.56$.}
\label{fig:F_4}
\end{figure}

\begin{figure}[tb]
\centering
\includegraphics[width=0.95\linewidth]{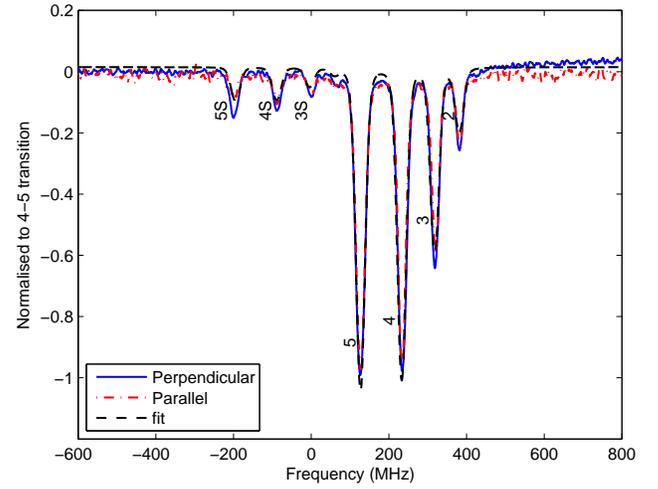}
\caption{Doppler free spectra with 852\,nm light locked on $6S_{1/2}(F=4)\rightarrow 6P_{3/2}(F^\prime=5)$ and 685\,nm light scanned across the $6S_{1/2}(F=3)$ to $5D_{5/2}(F^\prime=\mbox{2--5})$ transitions.
We use the same notation as in Fig.~\ref{fig:F_4}}
\label{fig:F_3}
\end{figure}

Figures~\ref{fig:F_4} and \ref{fig:F_3} shows two typical Doppler-free pump-probe spectra.
Here the 852\,nm laser is frequency locked to the $6S_{1/2}(F=4)\rightarrow 6P_{3/2}(F^\prime=5)$ transition whereas the 685\,nm laser is scanned across either the $6S_{1/2}(F=4)$ to $5D_{5/2}(F^\prime)$ multiplet for fig.~\ref{fig:F_4} or the $6S_{1/2}(F=3)$ to $5D_{5/2}(F^\prime)$ multiplet for fig.~\ref{fig:F_3}. These experimental spectra call for some remarks concerning selection rules, amplitudes, sideband resonances and linewidths.

\begin{enumerate}
\item The 685\,nm light couples the hyperfine transitions satisfying the quadrupole transition selection rules,
i.e.\ $-2\leq\Delta F \leq 2$.
In figure~\ref{fig:F_4}, excitation spectra on the $F=4\rightarrow F^\prime = 2,3,4,5\,\&\, 6$ lines are indeed observed.
Since the lasers at 685\,nm and 852\,nm are sharing the same $F=4$ ground state, the signal is revealed mainly through optical de-pumping of this state.
Thus the transmission signal at 852\,nm will be in-phase with the 18\,kHz modulation leading to a positive, i.e.\ emission-like spectrum, as seen in figure~\ref{fig:F_4}.
On the contrary, if the lasers are not sharing the same ground state, the transmission signal at 852\,nm will originate in an optical re-pumping scheme and be out-of-phase with the 18\,kHz modulation leading to a negative, i.e.\ absorption-like spectrum.
This situation is illustrated on figure~\ref{fig:F_3}.
Here the optical pumping from the ground state $F=3$ to the ground state $F=4$ is done through a 685\,nm absorption event followed by a $5D_{5/2}\rightarrow 6P_{3/2}\rightarrow 6S_{1/2}$ radiative cascade event.
Both events have the same $-2\leq \Delta F \leq 2$ selection rules.
Thus only the four $F=3 \rightarrow F^\prime =2,3,4\, \&\, 5$ lines of the E2 transition are observed, whereas the $F=3\rightarrow F^\prime=1$ transition is missing since it cannot decay back to the $6S_{1/2}\,F=4$ state.

\item
The different line intensities depend on the hyperfine levels involved in the optical pumping process, and eventually on the relative polarizations of the 685\,nm and 852\,nm lasers.
However, one should note that due to fast transit time in the beam, the 852\,nm irradiation does not \new{induce}\old{have enough time to produce} optical pumping between magnetic sublevels in the ground state.
Also, for a $F=4\rightarrow 5$ transition, a linearly polarized irradiation is nearly equivalent to an isotropic light probe.
This explains why the spectra are similar for parallel and perpendicular polarizations, as seen in figs~\ref{fig:F_4} and~\ref{fig:F_3}.
One may thus calculate the line intensities, based only on $S^Q_{FF^\prime}$ (\eqref{eqn:E2_str}), and E1 emission line strengths (for both $5D\rightarrow 6P$ and $6P\rightarrow 6S$ as given by:
\begin{equation}
S^D_{F^\prime F^{\prime \prime}}=(2F{''}+1)(2J' +1)\sj{J'}{J''}{1}{F''}{F'}{I}^2.
\end{equation}

Line amplitudes will depend on the de-population and re-population in the $6S$ level. The signal strength can be written as
\begin{equation}
\label{eqn:signal}
S(F_1,F,F_2)=\delta_{F_1,F_2}S^Q_{F_1 F} - \alpha \sum_{F''} S^Q_{F_1F}S^D_{FF''}S^D_{F'' F_2}\ ,
\end{equation}
where $\delta_{i,j}$ is the usual Kronecker delta.
The sum is over the $6P_{3/2}(F'')$ hyperfine levels (F indicates the pumped $5D_{5/2}$ hyperfine state) and
\begin{equation}
\alpha=\frac{\gamma^\mathrm{se}_{5D}}{\gamma_t+\gamma^\mathrm{se}_{5D}},
\end{equation}
is a re-pumping ratio which takes into account the transit relaxation rate,
described phenomenologically by an exponential decay rate $\gamma_t$.
This transit relaxation is mainly effective in the long-lived 5D state.
By increasing the 5D relaxation rate, it diminishes the re-pumping  back to the ground state~(second term in \eqref{eqn:signal}).
The transit relaxation rate may be approximated roughly by $\gamma_t \sim <v_\perp>/w_0$.
With the transverse mean thermal velocity across the laser beam, $<v_\perp> \approx 170$\, ms$^{-1}$, and the beam waist, $w_0 \approx 125$\,µm, we predict $\alpha \approx $ 0.4. This is a quite approximate prediction because transit time broadening is mainly governed by a Gaussian probability distribution.
If $F_1=F_2$ (same hyperfine ground state), the line intensity depends on both de-population and re-population pumping, as well as on the transit time relaxation.
As shown in figure~\ref{fig:F_4}, a very good match is obtained for $\alpha = 0.56\pm 0.02$ (i.e.\ $\gamma_t = 0.8 \gamma^{\mathrm{se}}_{5D}$) a value not far from the rough estimate above.
Note the major role of transit broadening. In its absence ($\alpha=1$), the repopulation in the $F=4$ ground state [after E2 excitation to the $5D_{5/2}(F=6)$ level] should exactly cancel the depopulation pumping (inside the close three-level system, $F=4-6-5-4$).
The observation of the $F=6$ resonance (Fig.~\ref{fig:F_4}) is a direct evidence for the transit time influence responsible for losses in the optical pumping process.
On the other hand, for the $F_1\neq F_2$ cross resonances (Fig.~\ref{fig:F_3}), there is no similar effect. Only the overall spectrum amplitude will depend on $\alpha$.

\item
Small amplitudes transmission peaks, appearing on the red side of the main resonances (Figs.~\ref{fig:fluo}--\ref{fig:F_3}) are sideband resonances. They correspond to the $6S_{1/2}(F=4)\rightarrow 6P_{3/2}(F=4)$ transition resonantly excited for a non-zero atomic velocity group\cite{feld_laser-induced_1969,ducloy_laser_1978}.
Taking into account the different Doppler shifts, we find a frequency shift between the two sets of resonances of $\vec{k}_{685}/\vec{k}_{852}\times\Delta_{5-4}=-312$\,MHz,where $\Delta_{5-4}=251$\,MHz is the $F=5-F=4$ hyperfine splitting of the $6P_{3/2}$ excited state.
Since the laser beams are counter propagating, the sidebands lie on the red side of the resonances.
With co-propagating beams, the shift would be on the blue side of the main resonances.

\item
The ultimate resonance linewidth is $\sim6.6$\,MHz.
The experimental linewidths are larger, about 15\,MHz. The extra broadening may be due to the 852\,nm power broadening, collisions and the 685\,nm laser jitter.
\end{enumerate}

In conclusion, we have observed Doppler-free spectral lines on the Cs $6\LQ{S}_{1/2}-5\LQ{D}_{5/2}$ transition and analyzed some of their properties, like selection rules and saturation intensities.
This work paves the way for further investigations like the  general polarization characteristics of E2 absorption, or the influence of near-field surface potentials on E2 line emission.
Angular momentum conservation in matter-light interaction involves photon spin only in Gaussian light beams.
In the work presented here, one thus expects to induce $\Delta M = \pm 1$ transitions in E2 absorption at 685\,nm.
If instead one uses focused Laguerre-Gauss (LG) beams, one expects that orbital angular momentum of light can be absorbed and thus produce $\Delta M= 0,\pm 1,\pm 2$ transitions\cite{klimov_detecting_2009,klimov_mapping_2012}.
These could be observed via E2 Zeeman transitions in applied magnetic fields.
E2 transitions can be monitored in zero-electric-field regions of LG beams and should allow one to map their spatial intensity distributions\cite{klimov_mapping_2012}.
Another prospect lies in the analysis of atom-metamaterial hybrid systems.
Engineering of atom-surface interactions in such hybrid devices have been recently demonstrated by monitoring the atom response on E1 transitions\cite{chan:2016}.
Similar work on E2 transitions will allow exploring e.m. field gradients in the vicinity of metamaterials.


{\bf FUNDING.} This work was supported by the Singapore Ministry of Education(501100001459) Academic Research Fund Tier 3 (Grant No. MOE2011-T3-1-005).


{\bf ACKNOWLEDGEMENT.} The authors thank A. Laliotis for work in initial setting up of the experiment.


\begin{thebibliography}{10}
\newcommand{\enquote}[1]{``#1''}

\bibitem{letokhov_nonlinear_1977}
V.~S. Letokhov and V.~P. Chebotayev, \emph{Nonlinear laser spectroscopy}
  (Springer-Verlag, 1977).

\bibitem{demtroder_laser_2003}
W.~Demtr\"{o}der, \emph{Laser spectroscopy : basic concepts and
  instrumentation}, Advanced texts in physics (Berlin ; New York : Springer, 2003).

\bibitem{tsigutkin_observation_2009}
K.~Tsigutkin, D.~Dounas-Frazer, A.~Family, J.~E. Stalnaker, V.~V. Yashchuk, and
  D.~Budker, Phys. Rev. Lett. \textbf{103}, 071601 (2009).

\bibitem{tamm_spectroscopy_2000}
C.~Tamm, D.~Engelke, and V.~B\"{u}hner, Phys. Rev. A \textbf{61}, 053405
  (2000).

\bibitem{oskay_single-atom_2006}
W.~H. Oskay, S.~A. Diddams, E.~A. Donley, T.~M. Fortier, T.~P. Heavner,
  L.~Hollberg, W.~M. Itano, S.~R. Jefferts, M.~J. Delaney, K.~Kim, F.~Levi,
  T.~E. Parker, and J.~C. Bergquist, Phys. Rev. Lett. \textbf{97}, 020801
  (2006).
\bibitem{rosenband_frequency_2008}
T.~Rosenband, D.~B. Hume, P.~O. Schmidt, C.~W. Chou, A.~Brusch, L.~Lorini, W.~H. Oskay, R.~E. Drullinger,
T.~M. Fortier, J.~E. Stalnaker, S.~A. Diddams, W.~C. Swann, N.~R. Newbury, W.~M. Itano, D.~J. Wineland, and J.~C. Bergquist
, Science \textbf{319}, 1808 (2008).

\bibitem{tojo_absorption_2004}
S.~Tojo, M.~Hasuo, and T.~Fujimoto, Phys. Rev. Lett. \textbf{92}, 053001
  (2004).

\bibitem{tojo_precision_2005}
S.~Tojo, T.~Fujimoto, and M.~Hasuo, Phys. Rev. A \textbf{71}, 012507 (2005).

\bibitem{weber_accurate_1987}
K.~H. Weber and C.~J. Sansonetti, Phys. Rev. A \textbf{35}, 4650 (1987).

\bibitem{ponciano-ojeda_observation_2015}
F.~Ponciano-Ojeda, S.~Hern\'andez-G\'omez, O.~L\'opez-Hern\'andez,
  C.~Mojica-Casique, R.~Col\'{\i}n-Rodr\'{\i}guez,
  F.~Ram\'{\i}rez-Mart\'{\i}nez, J.~Flores-Mijangos, D.~Sahag\'un,
  R.~J\'auregui, and J.~Jim\'enez-Mier, Phys. Rev. A \textbf{92}, 042511 (2015).

\bibitem{sargsyan_giant_2014}
A.~Sargsyan, A.~Tonoyan, G.~Hakhumyan, A.~Papoyan, E.~Mariotti, and
  D.~Sarkisyan, Laser Phys. Lett. \textbf{11}, 055701 (2014).


\bibitem{nez_optical_1993}
F.~Nez, F.~Biraben, R.~Felder, and Y.~Millerioux, Opt. Comm.
  \textbf{102}, 432 (1993).

\bibitem{kumar_precision_2014}
P.~V.~Kiran Kumar and M.~V. Suryanarayana, Pramana \textbf{83}, 189 (2014).

\bibitem{feld_laser-induced_1969}
M.~S. Feld and A.~Javan, Phys. Rev. \textbf{177}, 540 (1969).

\bibitem{ducloy_laser_1978}
M.~Ducloy, J.~R.~R. Leite, and M.~S. Feld, Phys. Rev. A \textbf{17}, 623
  (1978) and references therein.

\bibitem{thoumany_optical_2009}
P.~Thoumany, T.~Hänsch, G.~Stania, L.~Urbonas, and T.~Becker Opt. Lett.
  \textbf{34}, 1621 (2009).

\bibitem{klimov_detecting_2009}
V.~Klimov, D.~Bloch, M.~Ducloy, and J.~R. Rios~Leite, Opt. Exp. \textbf{17},
  9718 (2009).

\bibitem{klimov_mapping_2012}
V.~V. Klimov, D.~Bloch, M.~Ducloy, and J.~R. Rios~Leite, Phys. Rev. A
  \textbf{85}, 053834 (2012).

\bibitem{deguchi_simulation_2009}
K.~Deguchi, M.~Okuda, A.~Iwamae, H.~Nakamura, K.~Sawada, and M.~Hasuo, J. Phys. Soc. Jap. \textbf{78}, 024301 (2009).

\bibitem{kern_strong_2012}
A.~M. Kern and O.~J.~F. Martin, Phys. Rev. A \textbf{85}, 022501 (2012).

\bibitem{yannopapas_giant_2015}
V.~Yannopapas and E.~Paspalakis, J. Mod. Opt. \textbf{62}, 1435
  (2015).

\bibitem{diberardino_lifetime_1998}
D.~DiBerardino, C.~E. Tanner, and A.~Sieradzan, Phys. Rev. A \textbf{57},
  4204 (1998).

\bibitem{weissbluth_atoms_1978}
M.~Weissbluth, \emph{Atoms and molecules} (Academic Press, New York, 1978) Chapter 23.

\bibitem{chan:2016}
S.~A. Aljunid, E.~A. Chan, G.~Adamo, M.~Ducloy,  D.~Wilkowski, and N.~I. Zheludev, Nano Lett. \textbf{16}, 3137 (2016).


\end{thebibliography}
\end{document}